\documentclass[preprint,showpacs,preprintnumbers,amsmath,amssymb, showkeys,superscriptaddress,titlepage]{revtex4-2}
\usepackage{graphicx}
\usepackage{dcolumn}
\usepackage{bm}
\usepackage{hyperref}
\usepackage{textcomp}

\begin{document}
	
	\title{Exposure of Track Detectors in Xenon Ion Beams in NICA Accelerator Complex}
	
	\author{\firstname{A.A.}~\surname{Zaitsev}}
	\email{zaicev@jinr.ru}
	\affiliation{Joint Institute for Nuclear Research, Dubna, 141980, Russia}%
	\affiliation{P.N. Lebedev Physical Institute of the Russian Academy of Sciences, Moscow, 119991, Russia}
	\author{\firstname{P.I.}~\surname{Zarubin}}
	\affiliation{Joint Institute for Nuclear Research, Dubna, 141980, Russia}%
	\affiliation{P.N. Lebedev Physical Institute of the Russian Academy of Sciences, Moscow, 119991, Russia}
	
	\author{\firstname{S.D.}~\surname{Murashko}}
	\affiliation{International Sakharov Environmental Institute, Belarusian State University, Minsk, 220030, Belarus}%

	\author{\firstname{N.}~\surname{Marimuthu}}
	\affiliation{Joint Institute for Nuclear Research, Dubna, 141980, Russia}%
	
	\author{\firstname{A.A.}~\surname{Slivin}}
	\affiliation{Joint Institute for Nuclear Research, Dubna, 141980, Russia}%
	
	\author{\firstname{G.A.}~\surname{Filatov}}
	\affiliation{Joint Institute for Nuclear Research, Dubna, 141980, Russia}%
	 	
	
	\begin{abstract}
		The results of the analysis of solid-state track detectors CR39 and nuclear photoemulsion plates irradiated in beams of accelerated xenon ions with energies of 3.2 MeV/n and 3.8 GeV/n at the NICA accelerator complex are presented.
	\end{abstract}
	
	\maketitle
	
	\section{Introduction}
	
	At the fourth session of the NICA accelerator complex, beams of accelerated xenon ions were generated for the first time. It was proposed to use a relatively cheap and flexible method of film registration of charged particles for the purpose of photo-fact observation of accelerated beams, as well as profilometric measurements of the formed beams. Samples of the CR39 solid state track detector and a plate of nuclear photoemulsion have been irradiated in the beam injection area into the Booster (the SOCHI test station - the Chip Irradiation Station) and on the extracted beam from the Nuclotron at the F3 point and in the experimental zone of the BM@N (Baryonic Matter at Nuclotron) facility.
	
	\section{REGISTRATION OF CHARGED PARTICLES BY THE METHOD OF SOLID-STATE TRACK DETECTORS}
	
The solid-state track detector (SSTD) method is one of the methods of profilometry of beams of charged nuclear fragments, based on the use of solid-state materials capable of recording tracks of ions passing through them. When ions interact with a solid material, tracks are formed, the location and shape of which depend on the energy, type of ions and angle of incidence. By measuring the parameters of tracks on the surface of a solid-state detector, it is possible to obtain information on the spatial and charge distribution of ions in the cross section of a monoenergetic beam \cite{1,2}.

The SSTD method has some advantages over other profilometry methods, such as high spatial resolution, absence of detector dead time and registration efficiency. This justifies the possibility of using SSTD to monitor the density, position and integrated intensity of a beam of heavy relativistic ions. One of the SSTD samples is a plastic monomer allyl diglycol carbonate (ADC), which has the commercial name CR39 (Columbia Resin No. 39). CR39 is a transparent rigid plastic with a density of about 1.30 g/cm$^3$. This material is one of the most sensitive SSTDs and can directly register energetic protons, alpha particles and heavier nuclei \cite{1,2}.

A heavy ion passing through the CR39 material causes radiation damage to molecular compounds, thereby forming a latent track. To observe the tracks using optical microscopy, they must be enlarged by 
chemical etching of the detector. Typical chemical etchants for CR39 are NaOH and KOH solutions. The concentration and temperature of the etchant, as well as the etching time, are selected taking into account the irradiation performed (dependence of the energy and charge of the particles and their total flux through the sample). 
With 
chemical 
etching, 
the 
area 
of 
radiation damage to the detector 
material is destroyed faster than the undamaged area. As a result of erosion of the latent track, a conical etched pit appears in the detector material, which can be clearly distinguished under an optical microscope (Fig. 1).
	
	\begin{figure}
		\centering
		\includegraphics[width=0.5\textwidth]{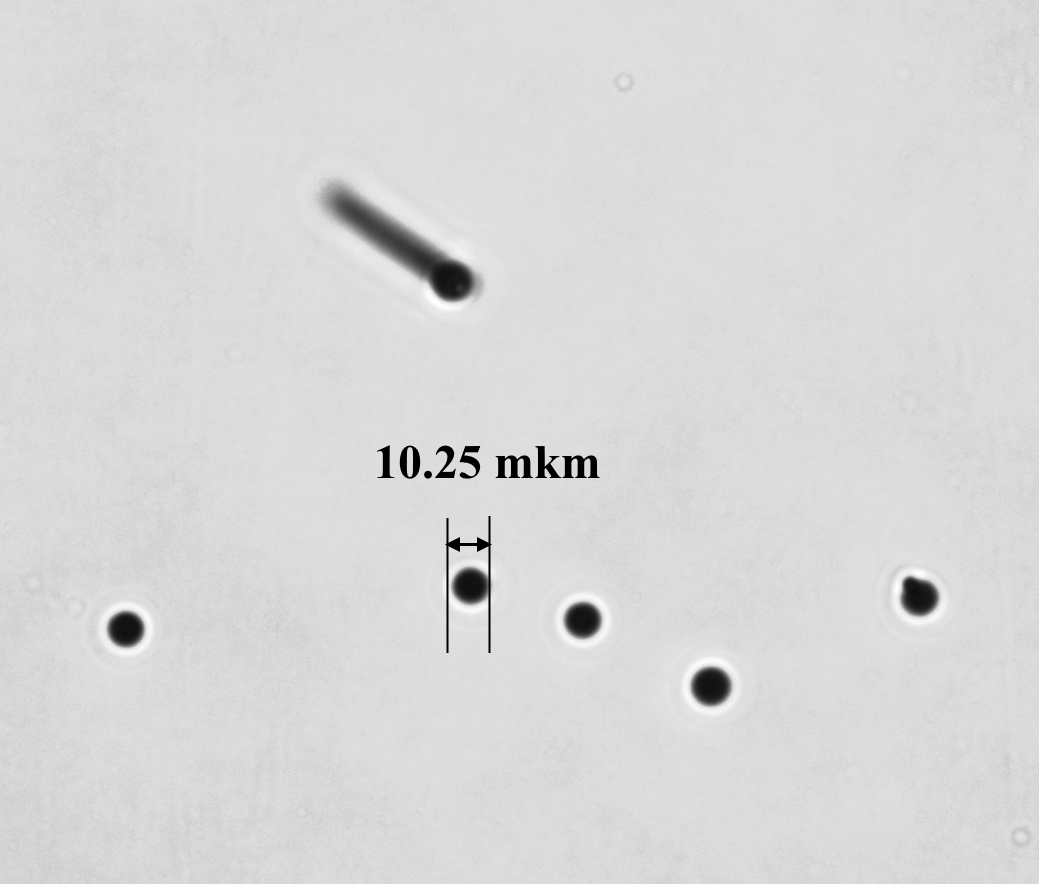}
		\caption{An enlarged image of the CR39 SSTD sample irradiated in a beam of $^{124}$Xe$^{+28}$ ions with an energy of 3.2 MeV/n. The lower part of the figure shows individual tracks of incoming ions at a right angle to the detector plane as black circles with dimensions of 10 $\mu$m. The upper track corresponds to an ion that entered at an angle to the detector plane.}
	\end{figure}
	
The sensitivity of track registration is determined by the relative etching rate of the damaged area and depends on the ionization energy losses of the particle in the material. For CR39 SSTD, the minimum linear energy transfer at which tracks are formed is about 3 keV/$\mu$m \cite{3}.
	
\section{ANALYSIS OF CR39 SOLID STATE TRACK DETECTORS}

The CR39 samples have been irradiated on the experimental setup in the experimental zone of the BM@N facility during the 4th commissioning session of the NICA complex. The energy of the $^{124}$Xe$^{+54}$ nuclear beam was 3.85 GeV/n, the integral detector load was 5 spills with a total flux of about 10$^6$ ions. The CR39 samples were located between the cathode-strip chamber and the calorimeter, with the normal to the CR39 surface coinciding with the direction of the primary beam. The CR39 detector under study was initially a 50x50 mm$^2$ plate with a thickness of 1 mm. After irradiation, the CR39 sample was etched in an aqueous solution with a NaOH concentration of 6 mol per liter. During etching, all samples were in a thermostat maintaining a temperature of (85$\pm$0.1) $^{\circ}$C. The etching time was 20 minutes.

The etched samples were scanned using a motorized Olympus BX63 microscope with the proprietary Olympus cellSens software, which provides automatic panoramic imaging with maps of precise positioning of the optical focus on the surface of the sample. This software automatically recognized track patterns on the panoramic image. During irradiation, the angle of incidence of Xe ions was close to perpendicular, resulting in the formation of tracks of approximately circular shape. To optimize the time of manual digitization, the surface was scanned with a 40$\times$ objective using the automatic stitching function of panoramic images using a focus map. The spatial distribution of the incoming tracks is shown in Fig. 2. The main region of the beam (core) has an elliptical shape with an ellipse tilt of about 45$^{\circ}$ in the XY plane. The beam ellipse, within which the track density exceeds 10$^3$ mm$^{-2}$, has the dimensions of the major and minor semi-axes of about 16 mm and 8 mm, respectively.

Tracking the position of the entry and exit of xenon nucleus tracks from the sample volume (1 mm thick) made it possible to calculate the beam rotation angle, which was $\Theta$ = 2.57$\pm$0.12 degrees relative to the normal to the sample surface.

\begin{figure}
	\centering
	\includegraphics[width=0.8\textwidth]{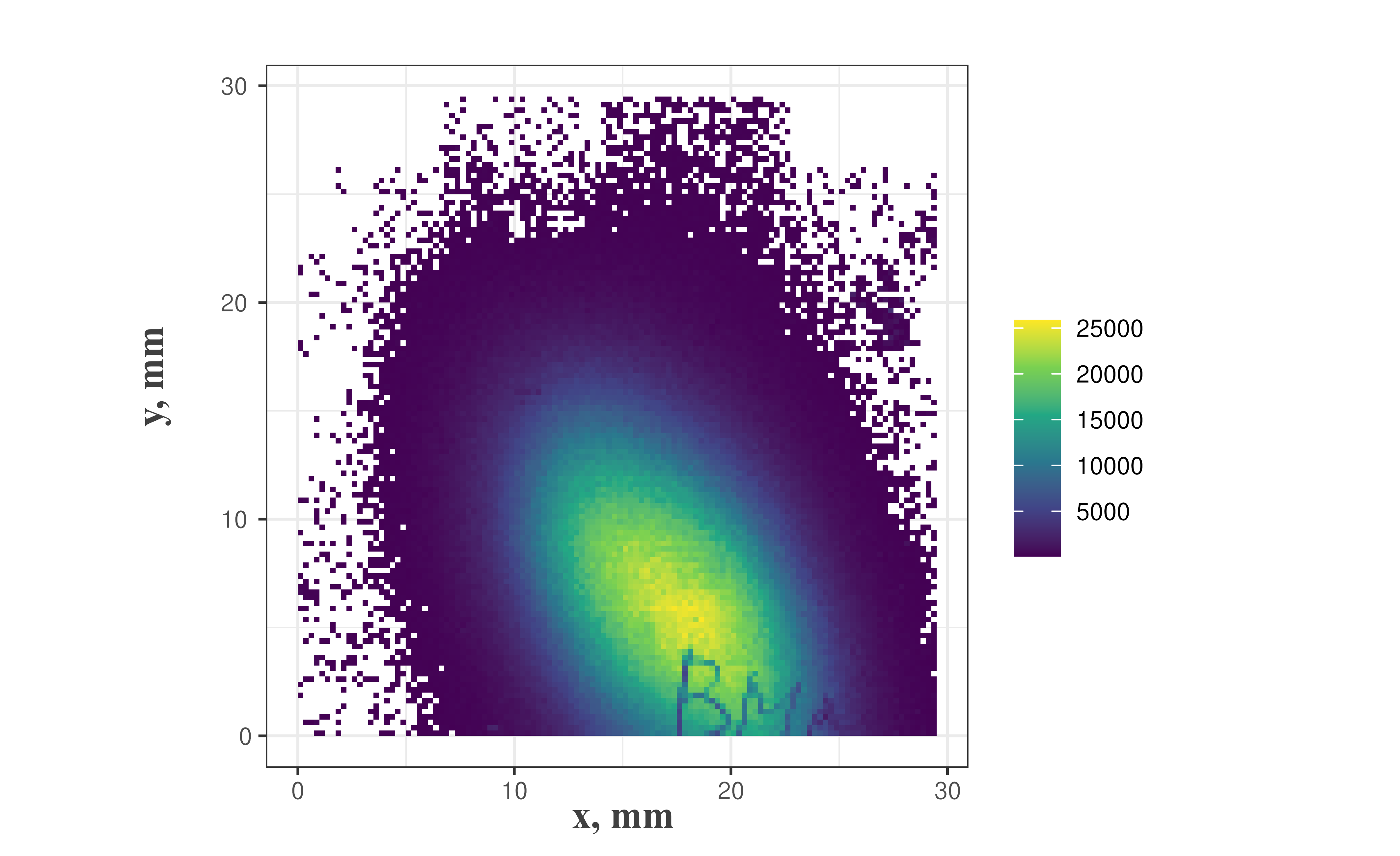}
	\caption{Spatial distribution of incoming xenon tracks in the CR39 SSTD sample irradiated in a beam of Xe nuclei. The "BMN" artifact is visible at the bottom of the figure, which is the marking on the surface of the SSTD sample. The color shows the intensity of the incoming Xe ion tracks in the scanned CR39 sample.}
\end{figure}

Due to the flexibility of the technique, it became possible to conduct experimental irradiation of CR39 plastic samples and PET (polyethylene terephthalate) films in low-energy xenon ion beams (3.2 MeV/n) in the vacuum chamber of the SOCHI station \cite{4}. The samples were irradiated under various conditions with a variable integral flux and beam focusing. Fig. 3 shows the distribution of incoming tracks into the CR39 SSTD sample in the case of a focused xenon ion beam in one discharge cycle.

\begin{figure}
	\centering
	\includegraphics[width=0.65\textwidth]{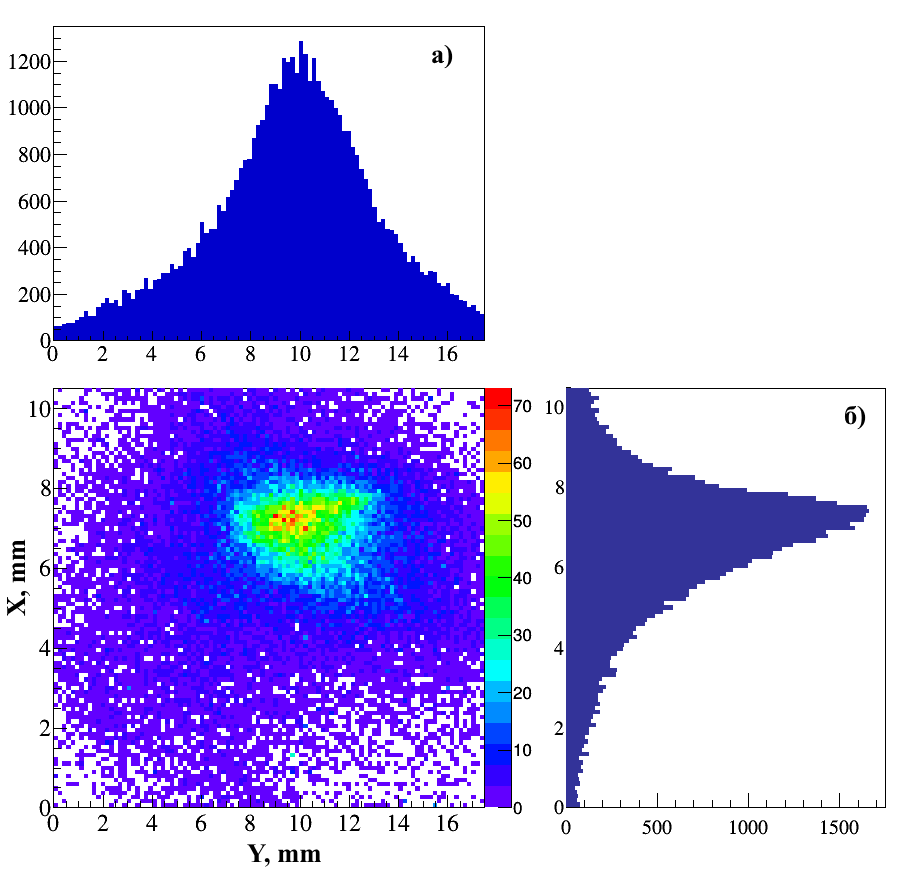}
	\caption{Distribution of incoming tracks of a focused beam of xenon ions over the surface of a CR39 sample. Inserts: (a) – projection onto the y-axis, (b) – projection onto the x-axis.}
\end{figure}

\section{IRRADIATION OF NUCLEAR PHOTOEMULSION PLATES IN XENON BEAMS WITH ENERGY OF 3.8 GeV/n}

The plates with nuclear photoemulsion (NTE) have been irradiated in the measuring pavilion at the F3 focus in the transport channel section (Nuclotron - 205 building). During irradiation of NTE stacks at the F3 focus, lenses 3K100 and 4K100 were switched off in order to uniformly load the NTE across the entire width. The orientation of the NTE stacks was chosen along the beam for longitudinal irradiation. The stacks consisted of 10 experimental layers. Each layer is a glass base measuring 9x12 cm$^2$ and 2 mm thick, onto which a sensitive NTE layer 100 $\mu$m thick was poured. The number of spill cycles was 1, 5 and 25 per stack with an integral particle flux per dump of $\approx$10$^6$. The developed plates are shown in Fig. 4. The stack with one dump cycle turned out to be optimal in terms of the density of Xe nuclei tracks in the NTE volume.

The second irradiation was carried out in the experimental zone of the BM@N facility \cite{5} between the cathode-strip chamber and the hadron calorimeter. The irradiation was carried out in the ``target OUT'' mode at the point where the beam passed, deflected by the SP41 dipole analyzing magnet (nominal field at the magnet pole is 8.5 kG). The irradiation was carried out in one cycle of dumping the beam of xenon nuclei with an energy of 3.8 GeV/n. Figure 5 shows an enlarged image of the NTE plate, where the tracks of the xenon nuclei are clearly visualized. Figure 6 shows an image of the peripheral interaction of the incident xenon nucleus with a heavy component from the NTE composition with multiple formation of secondary charged fragments of the projectile nucleus. Events of peripheral interactions of nuclei in NEE are identified by observation of a narrow stream of fragments oriented in the direction of the axis of the primary nucleus in the absence of tracks of fragments of the target nucleus and generated in the region of the collision vertex of nuclei. When visually viewing an irradiated plate, tracks of strongly ionizing relativistic particles are visible, accompanied by a plume of delta-electron tracks, which can be attributed to tracks of xenon.

\begin{figure}
	\centering
	\includegraphics[width=0.8\textwidth]{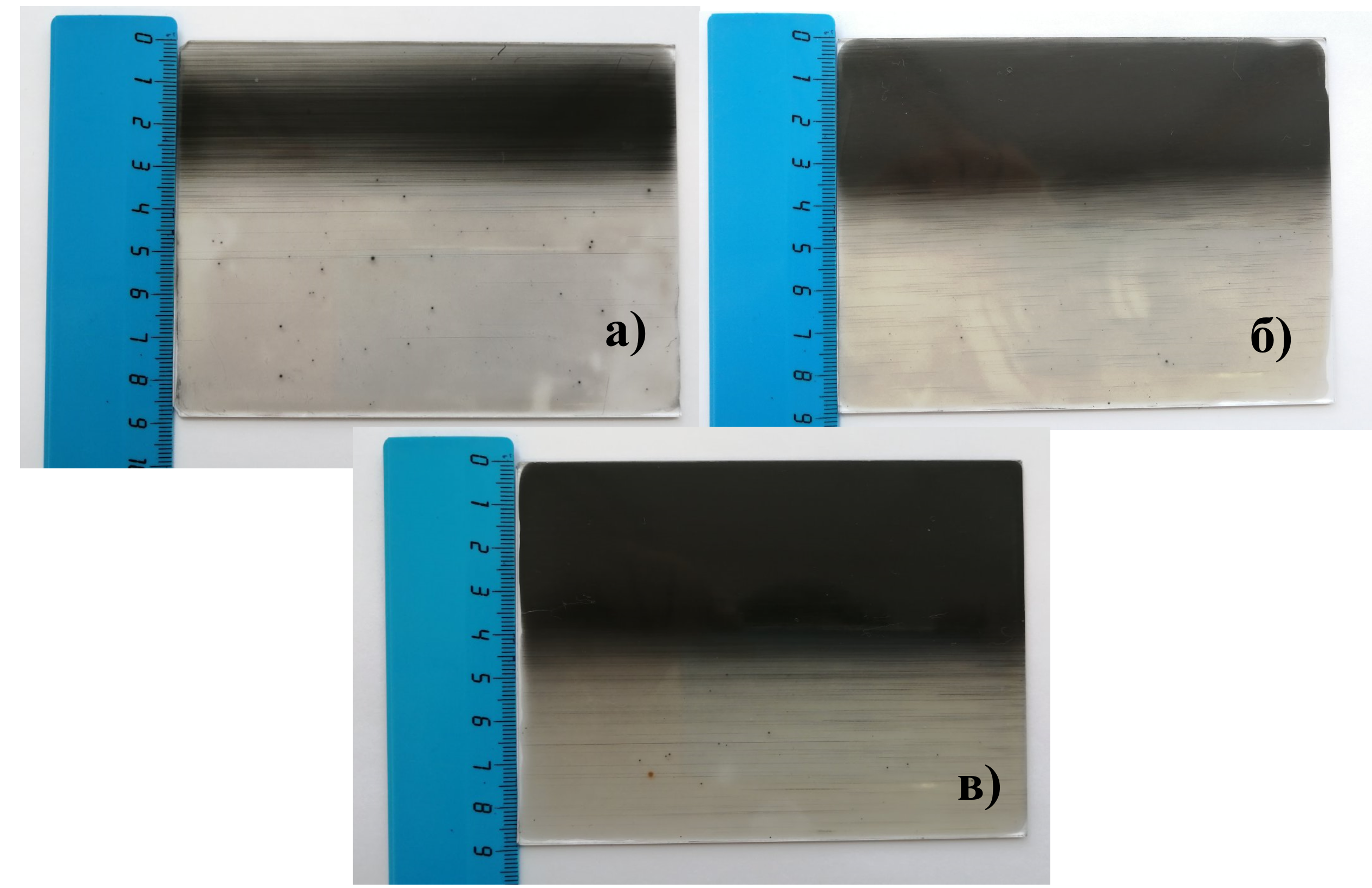}
	\caption{Image of irradiated NTE plates at F3 in 1 (a), 5 (b) and 25 (c) spills of discharges of a beam of accelerated xenon nuclei.}
\end{figure}

\begin{figure}
	\centering
	\includegraphics[width=0.8\textwidth]{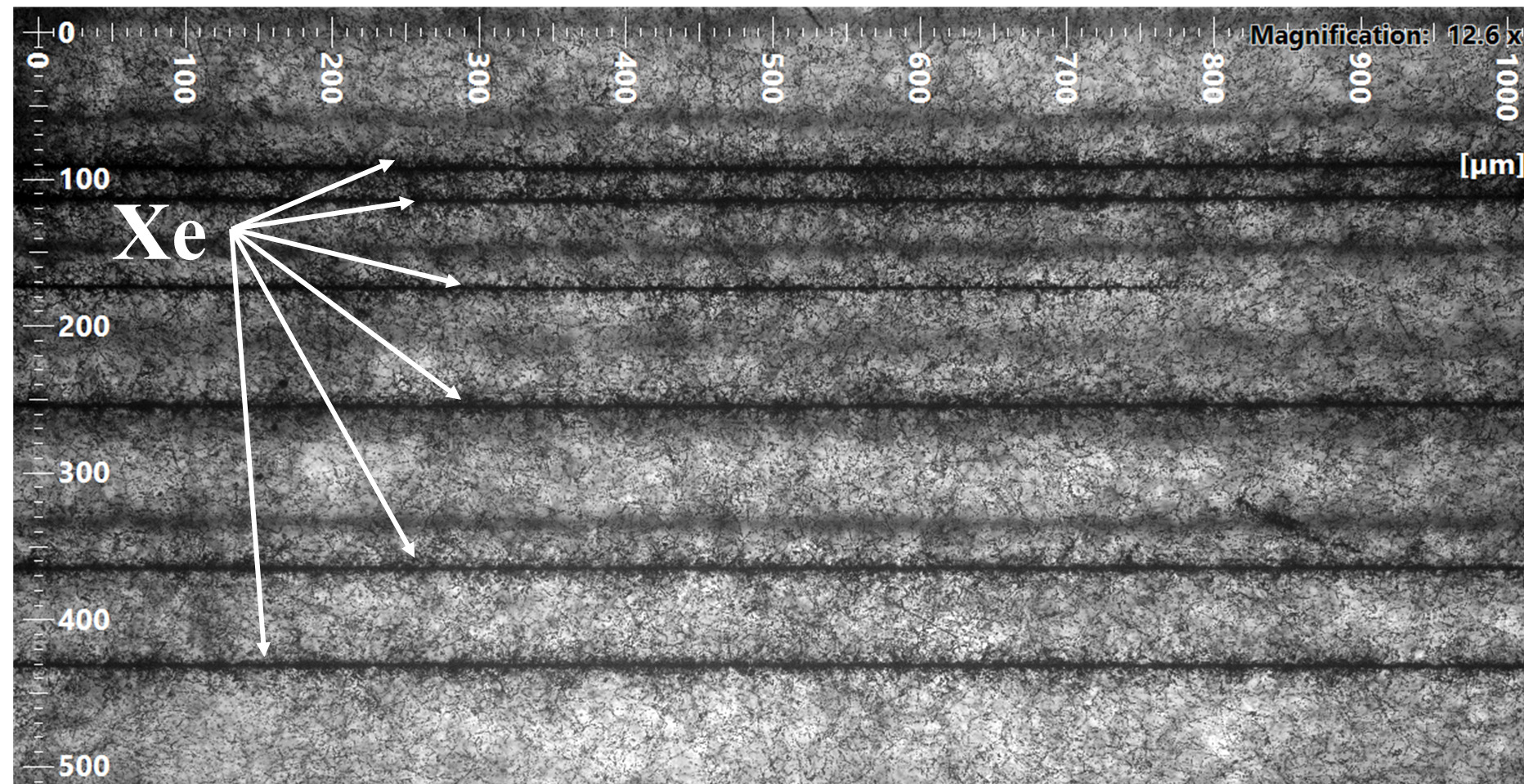}
	\caption{An enlarged image of a NTE layer irradiated in a beam of relativistic Xe nuclei at F3. The image clearly shows individual tracks of nuclei with a high density of formed $\delta$-electrons along the tracks. The XY scales are given for scale. The image was obtained on an Olympus BX63 motorized microscope with a 20$\times$ objective.}
\end{figure}

\begin{figure}
	\centering
	\includegraphics[width=0.8\textwidth]{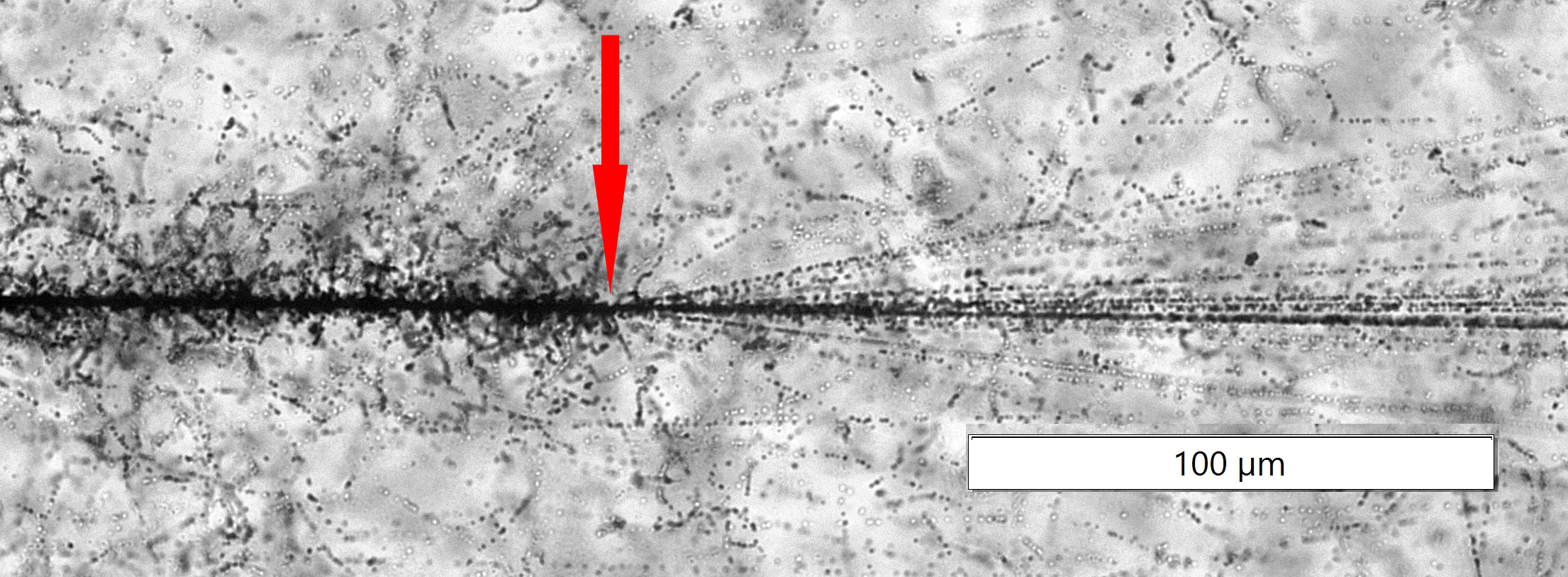}
	\caption{Macrophotograph of the peripheral interaction of a Xe nucleus in a NTE, obtained through a 40$\times$ objective of an Olympus BX63 microscope. The left part of the figure clearly shows the track of a xenon nucleus with a ``coat'' of $\delta$-electrons formed along the track. The arrow indicates the position of the vertex of the interaction of the incident nucleus with the subsequent formation of secondary fragments limited by a cone with an angle of 20$^{\circ}$.}
\end{figure}

\section{CONCLUSION}

Using the solid-state track detectors CR39, the profiles and intensities of low-energy beams of xenon ions ($^{124}$Xe$^{+28}$) extracted from the injector to the Chip Irradiation Station (SOCHI) have been reconstructed. Profilometric measurements of the extracted beams of relativistic xenon nuclei ($^{124}$Xe$^{+54}$) were carried out at the F3 point in the section of 
the transport channel and in the area of the BM@N experimental setup.
 The unique experimental
 material obtained in this way will subsequently form the basis for the analysis of multiple states of $\alpha$-particles and nucleons at the optimal energy of the incident nucleus within the BECQUEREL experiment (BEryllium Clustering QUEst in RELativistic multifragmentation \cite{6}).

\section{ACKNOWLEDGEMENTS}
The authors express their gratitude to Syresin E.M. for organizing and assisting in irradiating the CR39 SSTD samples in a xenon ion beam at the SOCHI station; to Piyadin S.M. and Anisimov S.Yu. for providing the opportunity to irradiate experimental layers of nuclear photoemulsion in beams of relativistic Xe nuclei.


\begin{thebibliography}{99}
		\bibitem{1} Picuz Jr. S.A., Skobelev I.Yu., Faenov A.Ya., Lavrinenko Ya.S., Belyaev V.S., Klyushnikov V.Yu., Matafonov A.P., Rusetsky A.S., Ryazantsev S. N. and Bakhmutova A.V. //Thermal physics of high temperatures. 2016. Vol. \textbf{54}. No. 3. P. 453-474.
		\bibitem{2} Jadrníčková I., Spurný F., Molokanov Aleksandr // Physics of Particles and Nuclei, Letters 2008. V. 5. P. 531-537.
		\bibitem{3} Kodaira S., Yasuda N., Kawashima H., Kurano M., Hasebe N., Doke T., Ota S. and Ogura K. // Radiation measurements 2009. V. \textbf{44}. No. 9-10. P. 861-864.
		\bibitem{4} G. A. Filatov 1, A. A. Slivin, E. M. Syresin, A. V. Butenko, A. S. Vorozhtsov, A. V. Agapov, K. N.Shipulin, S. Yu. Kolesnikov, V. N. Karpinsky, M. I. Kuznetsov, S. V. Kirov, A. V. Sergeev, A. R. Galimov, A. M. Tikhomirov, V. And . Tyulkin, D. S. Letkin, D. Oh. Leushin, A. V. Tuzikov // Physics of Particles and Nuclei, Letters 2022. Vol. \textbf{19}, No 5(244). J. 412-417.
		\bibitem{5} Mamaev M. (BM@N Collaboration) // Physics of Atomic Nuclei. 2023. V. \textbf{86}. No. 6. PP. 1346-1353.
		\bibitem{6} Web page of the BECQUEREL experiment: \href{http://becquerel.jinr.ru/}{becquerel.jinr.ru}.
		
	\end{thebibliography}
\end{document}